\documentclass[preprintnumbers, floatfix, preprintnumbers, letterpaper, superscriptaddress,nofootinbib, twocolumn]{revtex4}
\pdfoutput=1
\usepackage{graphicx}
\usepackage{microtype}
\usepackage{amsmath}
\usepackage{amssymb}
\usepackage{subfigure}
\usepackage{hyperref}
\usepackage{url}
\usepackage{xcolor}
\usepackage{color}
\usepackage{mathrsfs}
\usepackage{calrsfs}
\usepackage{amsfonts}
\usepackage{latexsym}
\usepackage{ragged2e}
\usepackage{epsfig}
\usepackage{textcomp}
\usepackage{phaistos}
\makeatletter
\renewcommand\@makefnmark{\hbox{\@textsuperscript{\normalfont\color{purple}\@thefnmark}}}
\renewcommand\@makefntext[1]{%
  \parindent 1em\noindent
            \hb@xt@1.8em{%
                \hss\@textsuperscript{\normalfont\@thefnmark}}#1}
\makeatother

\usepackage{caption}
\DeclareCaptionJustification{justified}{\leftskip=0pt \rightskip=0pt \parfillskip=0pt plus 1fil}
\definecolor{vividviolet}{rgb}{0.62, 0.0, 1.0}
\definecolor{amaranth}{rgb}{0.9, 0.17, 0.31}
\definecolor{palatinateblue}{rgb}{0.15, 0.23, 0.89}
\definecolor{brightpink}{rgb}{1.0, 0.0, 0.5}
\definecolor{cornflowerblue}{rgb}{0.39, 0.58, 0.93}
\definecolor{deepcarminepink}{rgb}{0.94, 0.19, 0.22}
\definecolor{radicalred}{rgb}{1.0, 0.21, 0.37}

\hypersetup{ linktoc=all,
    colorlinks, linkcolor={palatinateblue},
    citecolor={brightpink}, urlcolor={amaranth}
}

\graphicspath{{Images/}}

\graphicspath{{Images/}}



\def\sideremark#1{\ifvmode\leavevmode\fi\vadjust{\vbox to0pt{\vss
 \hbox to 0pt{\hskip\hsize\hskip1em
 \vbox{\hsize1.5cm\tiny\raggedright\pretolerance10000
 \noindent #1\hfill}\hss}\vbox to8pt{\vfil}\vss}}}%
\begin{document}

\title{Decoherence and Landauer's Principle in Qubit-Cavity \\
Quantum-Field-Theory Interaction}

\author{Hao Xu}
\email{haoxu@yzu.edu.cn}
\affiliation{Center for Gravitation and Cosmology, College of Physical Science and Technology, Yangzhou University, \\180 Siwangting Road, Yangzhou City, Jiangsu Province  225002, China}

\author{Si Yu Chen}
\affiliation{Center for Gravitation and Cosmology, College of Physical Science and Technology, Yangzhou University, \\180 Siwangting Road, Yangzhou City, Jiangsu Province  225002, China}

\author{Yen Chin \surname{Ong}}
\email{ycong@yzu.edu.cn}
\affiliation{Center for Gravitation and Cosmology, College of Physical Science and Technology, Yangzhou University, \\180 Siwangting Road, Yangzhou City, Jiangsu Province  225002, China}
\affiliation{Shanghai Frontier Science Center for Gravitational Wave Detection, Shanghai Jiao Tong University, Shanghai 200240, China}

\begin{abstract}
We consider quantum decoherence and Landauer’s principle in qubit-cavity quantum field theory (QFT) interaction, treating the qubit as the system and cavity QFT as the environment. In particular, we investigate the changes that occur in the system with a pure initial state and environment during the decoherence process, with or without energy dissipation, and compare the results with the case in which the initial state of the system is a mixed state and thus decoherence is absent. When we choose an interaction Hamiltonian such that the energy and coherence of the system change simultaneously, the population change of the system and the energy change are the same when the initial state is mixed. However, the decoherence terms increase the von Neumann entropy of the system. In this case the energy change and decoherence of the system are not independent physical processes. The decoherence process maintains unitarity. On the other hand, if the interaction Hamiltonian does not change the energy of the system, there is only the decoherence effect. The environment will be a distribution in the basis of the displaced number state and always increases the energy. Landauer's principle is satisfied in both cases.
\end{abstract}

\maketitle

\section{Introduction: Decoherence and Landauer's Principle}
The question of how a seemingly classical world can emerge from quantum mechanics is one of the greatest foundational problems in modern physics. In quantum mechanics, particles are described by a wave function satisfying the deterministic, linear Schrödinger equation
\begin{equation}
i\hbar\frac{d}{dt}|\psi\rangle=\hat{H}|\psi\rangle,
\end{equation}
therefore in principle we can compute the time evolution of the state given the initial state of the system and its Hamiltonian. However, the principle of superposition, which is the basic tenet of the quantum mechanics reflected in the linearity of the Schrödinger equation, predicts a probabilistic interpretation of our universe in macroscopic world, thus bringing a long and fascinating history of the measurement problem \cite{RevModPhys.76.1267}. After all, If the universe is governed by the fundamental principles of quantum mechanics at the microscopic level, why does the macro-world appear to be classical?

One of the reasons for this problem is our understanding was built on a tacit but incorrect assumption: the information about a system can be acquired without changing its state. This has led to the idealization of closed systems in physics. However, macroscopic systems are never isolated from their environments. In fact, a truly ideal closed system means that we cannot establish interactions with it in any way, and therefore cannot obtain any information about it. As emphasized by Zeh and Zurek, the Schrödinger equation applies only to a closed system, and the contribution from the environment must be included \cite{Zeh1970,Zurek1981,Zurek1982,Zurek2003}. The relevant theoretical framework for our study is therefore the theory of open quantum systems \cite{Breuer}.

This leads us to one of the keywords in the title of our work: decoherence. Decoherence, or more specifically, environment-induced decoherence, represents the loss of quantum coherence, a measure of the definite phase relation between different states of the system. In open quantum systems, the coupling to the environment defines and determines the physical properties of the system. Since almost every system is loosely coupled with other systems, decoherence can be viewed as the loss of information from one system to another. When we consider one of the systems in isolation, the evolution of the system is non-unitary, although the whole system is undergoing a unitary process. The process of decoherence is also the key component in explaining how the classical world emerges from the quantum regime. In the dynamical description of quantum-to-classical transition, the notion of ``classicality'' is understood as an emergent concept \cite{Schlosshauer}.

In many situations, we are used to treating the environment in quantum evolution as something whose details are unimportant and out of our control, whose quantum nature is often neglected. This is despite -- in principle -- the environment, regardless of its actual content, should also be treated quantum mechanically. There are some justifications for this approach. For example, the detailed dynamics of decoherence can be very complicated, and since interactions are almost ubiquitous, we do not even need to deliberately introduce the environment -- even cosmic rays and the microwave background radiation could lead to rapid decoherence of macroscopic bodies. Thus earlier studies have focused on aspects that are under our controls. However, since decoherence has become relevant, perhaps as the key obstacle, to practical applications such as quantum cryptography and quantum computation, it is necessary to further study the issue of the control of decoherence, including the backreaction effects brought to the environment.

In the present work we consider decoherence and Landauer’s principle in qubit-cavity quantum field theory (QFT) interaction. The qubit is arguably the simplest quantum system and the basis for quantum computing and quantum cryptography, whereas QFT is the description of Nature that contains all the fundamental interactions except for gravitation (whose quantization is still beyond our understanding). In particular, we will consider during the decoherence process of the system (qubit), with or without dissipation, the changes that occur in the environment (QFT). 

Landauer's principle is one of the bridges between thermodynamics and quantum information. It relates the entropy change of a system and the energy consumption of the environment, and it also provides a theoretical limit throughout the evolution \cite{landauer1961,landauer1996,Reeb2013}. Landauer's principle is based on the following four assumptions: (i) both the system $S$ and environment $E$ are described by Hilbert spaces, (ii) the environment is initially in a thermal state $\rho_E=e^{-\beta \hat{H}_E}/\text{Tr}(e^{-\beta \hat{H}_E})$, where $\hat{H}_E$ is the Hamiltonian of the environment and $\beta$ is the inverse temperature, (iii) system and environment are uncorrelated initially $\rho_{SE}=\rho_{S}\otimes\rho_{E}$, (iv) the process proceeds by unitary evolution $\rho'_{SE}=U\rho_{SE}U^{\dagger}$. If all the four assumptions are satisfied, Landauer's principle can be expressed in a form reminiscent of the first law of thermodynamics:
 \begin{equation}
\Delta Q\geqslant T_E\Delta S.
\label{bound}
\end{equation}
The quantity $\Delta Q :=\text{Tr}\left[\hat{H}_E(\rho'_E-\rho_E) \right]$ is the heat transferred to the environment $E$, where $\rho'_E$ and $\rho_E$ denote the final and initial state of $E$ respectively. Here $\Delta S := S(\rho_S)-S(\rho'_S)$ is the difference in the von Neumann entropy between the initial state $\rho_S$ and the final state $\rho'_S$ of the system $S$. The quantity $\rho_{S/E}:=\text{Tr}_{E/S}[\rho_{SE}]$ is the reduced density matrix.

Since the process of decoherence is inevitably accompanied by the change in von Neumann entropy, it is natural to ask what changes in the environment will occur accordingly. In addition, the process of decoherence may also be accompanied (or not accompanied) by energy changes in the system $S$, so we are interested in how this might affect the evolution of the whole system. In the present work we will answer these questions by analyzing the different forms of interactions between qubit and cavity QFT. Henceforth we shall adopt the natural unit system, setting $c=\hbar=k_B=1$ in all the analytical calculations and numerical analyses.

\section{{Analyses of the Models}}

In the qubit-cavity QFT interaction the total Hamiltonian $\hat{H}_{\text{total}}$ describing our system consists of three terms: $\hat{H}_{\text{total}}=\hat{H}_S+\hat{H}_E+\hat{H}_{\text{int}}$. The first term $\hat{H}_S$ is the free Hamiltonian of the system $S$ and in our case it is just a qubit so we can choose $\hat{H}_S=\frac{\Omega}{2}{\sigma_z}$, where $\sigma_z$ is the Pauli matrix. The second term $\hat{H}_E=\sum_{j=1}^{\infty}\omega_j a^{\dag}_ja_j$ is the free Hamiltonian of the cavity QFT, and $\hat{H}_{\text{int}}=\lambda \eta(\tau)m \phi[x(\tau)]$ is the interaction Hamiltonian, in which $\lambda$ is a weak coupling constant, and $\tau$ denotes the proper time. Here $\eta(\tau)$ is the so-called ``switching function'' that controls the interaction, $m$ is the monopole moment of the qubit, and $\phi[x(\tau)]$ is the field operator at the position of the qubit in the cavity. 

Before continuing the analysis we give a short discussion of the interaction Hamiltonian $\hat{H}_{\text{int}}$ to gain some intuition. In interaction picture the field operator $\phi[x(\tau)]$ reads
\begin{align}
\phi[x(\tau)] = \sum_{j=1}^{\infty}\left( a_je^{-i\omega_j t(\tau)} u_j\left[x(\tau)\right]+a^{\dagger}_je^{i\omega_j t(\tau)}u_j^*\left[x(\tau)\right] \right),
\label{int}
\end{align}
where the expression of $u_j\left[x(\tau)\right]$ depends on the boundary conditions of the cavity. Thus we can conclude that $\hat{H}_E$ and ${\hat{H}_{\text{int}}}$ do not commute with each other, regardless of the other quantities in ${\hat{H}_{\text{int}}}$. The energy expectation value of the cavity QFT is not a conserved quantity in the interaction. 

On the other hand, the monopole moment $m$ should be the linear combination of the Pauli matrices $\sigma_i$ $(i=x,y,z)$. For example, if we choose $m=\sigma_x$(the case $\sigma_y$ is similar), then in the interaction picture $m(\tau)=\sigma^{+}e^{i\Omega \tau}+\sigma^{-}e^{-i\Omega \tau}$, where $\sigma^{+}|0\rangle=|1\rangle$, $\sigma^{-}|1\rangle=|0\rangle$, where $|0\rangle$, $|1\rangle$ are the ground and excited states of the qubit, respectively. This means that the populations of the qubit density matrix can be exchanged between the ground and excited states, thus causing the change in energy. We can also obtain this by noticing that $[\hat{H}_S,\hat{H}_{\text{int}}]\neq 0$. If we have chosen $m=\sigma_z$ instead, we would have $[\hat{H}_S,\hat{H}_{\text{int}}]= 0$, and consequently the qubit does not undergo energy changes. In the present work we will discuss these two cases separately. 

Since Landauer's principle requires the initial state of the environment to be thermal, the density matrix for $E$ takes the form \cite{Olivares2012}
\begin{equation}
\rho_E=\bigotimes_{j=1}^{\infty}\sum_{n_j=0}^{\infty}\frac{\bar{n}_j^{n_j}}{(1+\bar{n}_j)^{1+n_j}}|n_j\rangle\langle n_j|,
\label{thermal}
\end{equation}
where for each integral value of $j$, $n_j \in [0,\infty)$, and $\bar{n}_j:={1}/{\left(e^{\frac{\omega_j}{T_E}}-1\right)}$ is the average photon number. For the qubit, we choose its initial state to be $|\psi_0\rangle=\sqrt{1-p}|0\rangle+\sqrt{p}|1\rangle$, in which both $\sqrt{1-p}$ and $\sqrt{p}$ are chosen to be real numbers. Thus the initial density matrix of the qubit is given by
\begin{equation}
\rho_S=\begin{pmatrix} 
    p & \sqrt{p(1-p)} \\

    \sqrt{p(1-p)} & 1-p
\end{pmatrix}.
\end{equation}
Note that this density matrix differs from $(1-p)|0\rangle\langle 0|+p|1\rangle\langle 1|$. Although both density matrices correspond to the same probability distribution, for the case of $(1-p)|0\rangle\langle 0|+p|1\rangle\langle 1|$, the population may be interpreted as classical probabilities, and the density matrix is not coherent. The main problem studied in this paper is the effect of the changes of the off-diagonal terms of the qubit on the whole system, while the diagonal terms of the qubit change or remain unchanged in the interaction.

\subsection{The case of $m=\sigma_x$}
In this case we assume that the coupling constant $\lambda$ is small, thus allowing the use of perturbation methods. The time evolution operator from time $\tau=0$ to $\tau=T$ is given by the Dyson series \cite{1209.4948}
\begin{align}
\hat{U}(T,0)=&\openone\underbrace{-i\int^{T}_{0}d\tau \hat{H}_{\text{int}}(\tau)}_{\hat{U}^{(1)}} \\ \notag
&\underbrace{+(-i)^2\int^{T}_{0}d\tau \int^{\tau}_{0}d\tau' \hat{H}_{\text{int}}(\tau)\hat{H}_{\text{int}}(\tau')}_{\hat{U}^{(2)}}+ ...\\ \notag
&\underbrace{+(-i)^n\int^{T}_{0}d\tau ... \int^{\tau^{(n-1)}}_{0}d\tau^{(n)} \hat{H}_{\text{int}}(\tau) ... \hat{H}_{\text{int}}(\tau^{(n)})}_{\hat{U}^{(n)}},
\label{dyson}
\end{align}
so the density matrix at a time $\tau=T$ is
\begin{equation}
\rho_{T}\!=\!\big[\openone+\hat{U}^{(1)}+\hat{U}^{(2)}+\mathcal{O}(\lambda^3)\big]\rho_0\big[\openone+\hat{U}^{(1)}+\hat{U}^{(2)}+\mathcal{O}(\lambda^3)\big]^{\dagger}.
\end{equation}
We can expand $\rho_T$ order by order as
\begin{equation}
\rho_{T}=\rho^{(0)}_{T}+\rho^{(1)}_{T}+\rho^{(2)}_{T}+\mathcal{O}(\lambda^3),
\end{equation}
where
\begin{align}
\rho^{(0)}_{T}&=\rho_0, \\
\rho^{(1)}_{T}&=\hat{U}^{(1)}\rho_0+\rho_0 \hat{U}^{(1)\dagger}, \\
\rho^{(2)}_{T}&=\hat{U}^{(1)}\rho_0 \hat{U}^{(1)\dagger}+\hat{U}^{(2)}\rho_0+\rho_0 \hat{U}^{(2)\dagger}.
\label{rho}
\end{align}

Let us first consider the $\rho^{(1)}_{T}$ term, in which $\hat{U}^{(1)}$ can be written as 
\begin{equation}
\hat{U}^{(1)}=\frac{\lambda}{i}\sum_{j=1}^{\infty}[\sigma^+ a_j^{\dagger}I_{+,j}+\sigma^- a_jI_{+,j}^*+\sigma^- a_j^{\dagger}I_{-,j}+\sigma^+ a_jI_{-,j}^*],
\label{U1}
\end{equation}
where $I_{\pm,j}$ is defined as
\begin{equation}
I_{\pm,j}:=\int^{T}_0 d\tau~e^{i\left[\pm \Omega \tau+\omega_jt(\tau)\right]} u_j\left[x(\tau)\right].
\label{I}
\end{equation}
Here we have already set $\eta(\tau)=1$ for $0\leqslant \tau \leqslant T$. Since our initial state \eqref{thermal} is thermal and contains only the diagonal terms, the operators $a_j$ and $a^{\dagger}_j$ from $\hat{U}^{(1)}$ acting on $\rho_0$ would only give rise to $|n_j-1\rangle\langle n_j|$ and $|n_j+1\rangle\langle n_j|$ ($\rho_0 \hat{U}^{(1)\dagger}$ is also similar). Since they are off-diagonal terms, when we subsequently perform the partial trace over the basis of cavity QFT to obtain the reduced density matrix of the qubit, we would necessarily obtain zero. This means that qubit-cavity QFT interaction has no effect at the $\lambda$ order. In the language of QFT, it is just the one point function $\langle \phi(x) \rangle=0$ for the thermal state. We can also easily see that the contributions from all odd-order $\lambda^{2n-1}~(n=1,2,3...)$ vanish. However, we need to emphasize that this is not a general result; it depends on the initial state of the QFT. If the initial state already contains some off-diagonal terms, such as in the case of coherent state, the odd-order $\lambda^{2n-1}$ interaction may create some diagonal terms and the $\lambda$ order would play the leading role. In the present work we are considering weak coupling and initial state of $E$ to be diagonal, so we shall restrict to at most the $\lambda^2$ terms and omit higher order ones.

Next we consider the $\lambda^2$ order term $\rho^{(2)}_{T}$. For the term $\hat{U}^{(1)}\rho_0 \hat{U}^{(1)\dagger}$, we already know from \eqref{U1} there are four terms in $\hat{U}^{(1)}$, so $\hat{U}^{(1)}\rho_0 \hat{U}^{(1)\dagger}$ would have sixteen terms in total. All we need to do is write out all the terms and pick the ones that are not zero after carrying out the partial trace. For example, we can have $|n_j+1\rangle\langle n_j+1|$ and $|n_j-1\rangle\langle n_j-1|$ in the field part, thus performing the partial trace would yield a non-vanishing result, so we can obtain the correction of the qubit density matrix. On the other hand, each term also contains $|I_{-,j}|^2$, $|I_{+,j}|^2$, etc. These quantities play the role of two-point function of the cavity QFT. As we previously mentioned, $u_j\left[x(\tau)\right]$ depends on the boundary condition. In this work we choose Dirichlet boundary condition and set $u_j\left[x(\tau)\right]\sim \sin[k_n x(\tau)]$. If the qubit is at rest $x(\tau)=\text{constant}$, then $t=\tau$ and $u_j\left[x(\tau)\right]$ gives a constant value. For $I_{-,j}$, as $\Omega=\omega_j$, the integrand will just be the value of $u_j\left[x(\tau)\right]$ at $x(\tau)=\text{constant}$, thus $I_{-,j}$ is proportional to the time $T$. In other words, the qubit and the cavity QFT resonate at the frequency $\Omega$ (we refer to this as the ``resonance effect''). However, for $\Omega \neq\omega_j$ or the case of $I_{+,j}$, the integration gives $\frac{1-e^{i(\pm \Omega+\omega_j)T}}{\pm \Omega+\omega_j}u_j(x)$, and the contributions from \eqref{I} becomes smaller and quickly decays for larger values of $\pm \Omega+\omega_j$. An analogous phenomenon in classical mechanics was reported in \cite{Smith2008}. In summary, we will select only those terms in $\hat{U}^{(1)}\rho_0 \hat{U}^{(1)\dagger}$ which are non-vanishing after partial tracing and furthermore must contain $|I_{-,j}|^2$.

Similarly, for both $\hat{U}^{(2)}\rho_0$ and $\rho_0 \hat{U}^{(2)\dagger}$, which each contain sixteen terms; we will pick the terms in the same manner. After some tedious calculation, by taking partial trace over the basis of cavity QFT we have the reduced density matrix of the qubit as 
\begin{equation}
\rho'_S=\begin{pmatrix} 
    p & \sqrt{p(1-p)} \\

    \sqrt{p(1-p)} & 1-p
\end{pmatrix}+
\begin{pmatrix} 
    \delta p & -\delta d \\

    -\delta d & -\delta p
\end{pmatrix},
\label{rhos2}
\end{equation}
where
\begin{align}
\delta p =\sum_{j=1}^{\infty}\lambda^2\big[ \left(\bar{n}_j(1-p)-(\bar{n}_j+1)p\right)|I_{-,j}|^2 \big],
\label{dp}
\end{align}
and 
\begin{align}
\delta d =\sum_{j=1}^{\infty}\lambda^2\sqrt{p(1-p)}\left(\bar{n}_j+\frac{1}{2}\right)|I_{-,j}|^2.
\label{dd}
\end{align}

Similarly, for the cavity QFT, partial tracing over the basis of the qubit yields
\begin{align}
&\text{Tr}_S(\hat{U}^{(1)}\rho_0 \hat{U}^{(1)\dagger})=\lambda^2\sum_{j=1}^{\infty} \bigg\{\big[p|I_{-,j}|^2 \big]
\\ \notag
&\times \sum_{n_j=0}^{\infty}\frac{\bar{n}_j^{n_j}(1+n_j)}{(1+\bar{n}_j)^{n_j+1}}|n_j+1\rangle\langle n_j+1|+\big(1-p)|I_{-,j}|^2 
\\ \notag
& \times \sum_{n_j=1}^{\infty}\frac{\bar{n}_j^{n_j}n_j}{(\bar{n}_j+1)^{n_j+1}}|n_j-1\rangle\langle n_j-1| \bigg\},
\end{align}
and
\begin{align}
&\text{Tr}_S(\hat{U}^{(2)}\rho_0)=\text{Tr}_S(\rho_0 \hat{U}^{(2)\dagger}) =-\frac{\lambda^2}{2}\sum_{j=1}^{\infty}\sum_{n_j=0}^{\infty} \Big[\big(n_j(1-p)
\\ \notag
&+(n_j+1)p\big)|I_{-,j}|^2\Big]\times \frac{\bar{n}_j^{n_j}}{(1+\bar{n}_j)^{1+n_j}}|n_j\rangle\langle n_j|,
\end{align}
where we have ignored the off-diagonal terms in $E$, because the energy of the free cavity QFT is only related to the diagonal terms.

If we compare the results here with the case of mixed state $(1-p)|0\rangle\langle 0|+p|1\rangle\langle 1|$ being the initial state in \cite{Xu:2021buk}, where the qubit evolution takes the form of
\begin{equation}
\begin{pmatrix} 
    p & 0 \\

    0 & 1-p
\end{pmatrix}\rightarrow \begin{pmatrix} 
    p+\delta p & 0 \\

    0 & 1-p-\delta p
\end{pmatrix},
\label{process}
\end{equation}
we find that the change of the cavity QFT and $\delta p$ of the qubit are the same in both cases, but the the off-diagonal term $\delta d$ in \eqref{rhos2} is new. This may be confusing for the coherence in qubit does not affect the $\Delta Q$. The reason here lies in the form we have chosen for the interaction Hamiltonian. In order to obtain $\Delta Q$, we need to know the reduced density matrix of the $E$ by taking partial trace over the basis of qubit. Taking $\hat{U}^{(1)}\rho_0 \hat{U}^{(1)\dagger}$ for example,  the matrix associated with the qubit can be written roughly as $m(\tau)\rho_0m(\tau')$. By some algebraic calculations we will find that the off-diagonal terms in the initial state of the qubit does not affect the trace of the matrix. Similarly we can also examine $\hat{U}^{(2)}\rho_0$ and $\rho_0 \hat{U}^{(2)\dagger}$, and we can also find the off-diagonal terms does not affect the trace. However, the off-diagonal terms of the initial state and their variations do affect the von Neumann entropy of the qubit and are thus relevant for Landauer's principle.

The first thing we can observe is that $\delta d$ is non-negative, which means that quantum coherence can only decrease during the interaction. According to the analysis of $|I_{-,j}|^2$, we know that the decoherence process will always continue, dominated by the resonance effect. Furthermore, since the initial state of the qubit is pure, the decoherence process must increase the von Neumann entropy. Diagonalizing $\rho'_S$ we have the von Neumann entropy of the final state
\begin{align}
S(\rho'_S)=-p_+\ln{p_+}-p_-\ln{p_-},
\label{S}
\end{align}
where 
\begin{align}
p_{\pm}&=\frac{1\pm \sqrt{1+(8p-4)\delta p-8\sqrt{p(1-p)}\delta d+4\delta d^2+4\delta p^2}}{2}
\\ \notag
       &= \frac{1}{2}\pm \frac{1}{2}(1+(4p-2)\delta p-4\sqrt{p(1-p)}\delta d)+\mathcal{O}(\delta p^2, \delta d^2)
\end{align}
is the eigenvalues of the $\rho'_S$. Omitting the higher-order terms $\mathcal{O}(\delta p^2, \delta d^2)$, we get
\begin{align}
p_{+}&=1-\left((1-2p)\delta p+2\sqrt{p(1-p)}\delta d\right),
\\ \notag
p_{-}&=(1-2p)\delta p+2\sqrt{p(1-p)}\delta d.
\label{popu}
\end{align}
Although $\delta p$ could be negative, in turns out that \emph{$p_{-}$ is always non-negative}. To see this, we simply insert the expressions \eqref{dp} and \eqref{dd} into the above formula of $p_{-}$, we have
\begin{align}
p_{-}=\sum_{j=1}^{\infty}\lambda^2 \left[ (2\bar{n}_j+1)p^2-2\bar{n}_j p+   \bar{n}_j   \right]|I_{-,j}|^2.
\end{align}
We can easily verify that this is indeed non-negative (the terms in the square bracket is a quadratic in $p^2$, with negative discriminant). This also means that in this case the change of energy and decoherence are \emph{not independent} physical processes. The energy change must be accompanied by the decoherence process to maintain unitarity.

Next we analyze the change in the von Neumann entropy. In Sec.II(B) of \cite{Xu:2021buk}, Landauer's principle $\Delta Q\geqslant T_E\Delta S$ has been verified for the qubit with the mixed initial state $(1-p)|0\rangle\langle 0|+p|1\rangle\langle 1|$, and since we have the same $\Delta Q$ and $\delta p$, if we can prove the entropy change in our case is always smaller than the case of \eqref{process}, Landauer's principle is naturally satisfied. We will find that this is indeed the case. For a \emph{non-vanishing} $p$, the entropy change of process \eqref{process} is 
\begin{align}
\Delta S = -\lambda^2\ln{\frac{1-p}{p}}\delta p.
\end{align}
On the other hand, if $p=0$, the von Neumann entropy will be
\begin{align}
S(\rho'_S) &= -(1-\delta p)\ln{(1-\delta p)}-\delta p\ln{\delta p}
\\ \notag
           &\approx -\delta p\ln{\delta p}+\delta p.
\end{align}
This is exactly our case, in which $p_-$ plays the role of $\delta p$. The derivative of the function $-(1- p)\ln{(1- p)}- p\ln{p}$ increases when $p$ gets smaller in the range $0<p<0.5$, and it becomes divergent in the limit $p\rightarrow 0$. Thus in our case the eigenvalues $p_{\pm}$ always give a larger von Neumann entropy and thus yield a smaller and negative $\Delta S$. When $p=0$, the two cases coincide. 

\begin{figure}
\begin{center}
\includegraphics[width=0.43\textwidth]{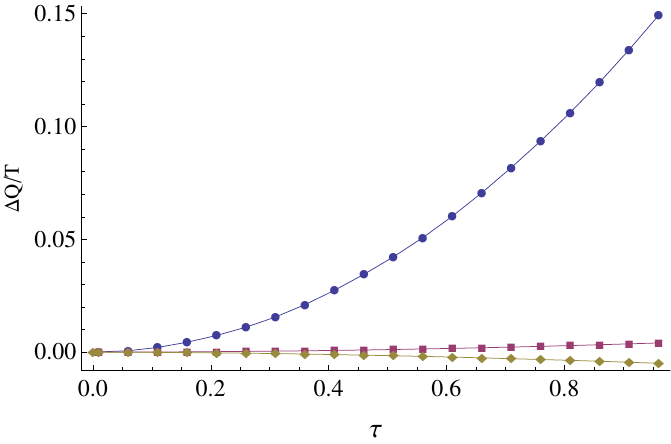}
\includegraphics[width=0.445\textwidth]{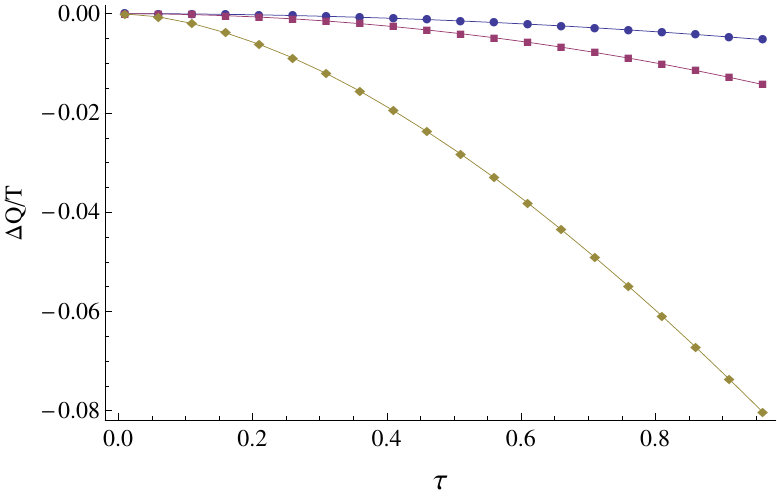}
\caption{The cases of $T_E=1$ and $T_E=100$ in single-mode approximation. In both figures we set $L=1.234$, $\Omega=\omega_{20}$, $p=0.2$, $x=0.52345$ in natural units. In each figure from top to bottom the curves correspond to ${\Delta Q}/{T_E}$, $\Delta S$ in process \eqref{process}, and $\Delta S$ in our case with decoherence, respectively.}
\label{fig1}
\end{center}
\end{figure}

In Fig.\ref{fig1} we present explicit examples of the quantities	${\Delta Q}/{T_E}$ and $\Delta S$ for both cases to illustrate this result. The two figures corresponds to $T_E=1$ and $T_E=100$ respectively. Here we apply the single-mode approximation, focusing on the resonance effect. In each figure from top to bottom the curves correspond to ${\Delta Q}/{T_E}$, $\Delta S$ in process \eqref{process}, and $\Delta S$ in our case with decoherence, respectively. For the case of $T_E=1$, both ${\Delta Q}/{T_E}$ and $\Delta S$ in process \eqref{process} are positive, while in the decoherence case $\Delta S$ is negative. For the case of $T_E=100$, ${\Delta Q}/{T_E}$ and $\Delta S$ are both negative, but  $\Delta S$ in the decoherence case is the smallest, which verifies the validity of Landauer's principle.

The case of $m=\sigma_y$ is similar to $m=\sigma_x$. It can also produce the variation in populations and can trigger decoherence. The calculation and conclusions are the same as $m=\sigma_x$, so we will not discuss it here.

\subsection{The case of $m=\sigma_z$}

The case of $m=\sigma_z$ is different from the $m=\sigma_x$ case, because with $[\hat{H}_S,\hat{H}_{\text{int}}]= 0$, the qubit does not undergo energy changes, and since $\sigma_z$ remains unchanged in the interaction picture, we also do not have the $e^{\pm i\Omega \tau}$ term. This is why there is no resonance effect in this case. The commutator of $\hat{H}_{\text{int}}$ at different times is not an operator, but just a c-number \cite{Hornberger2009}
\begin{equation}
[\hat{H}_{\text{int}}(\tau),\hat{H}_{\text{int}}(\tau')]=2i\sum_j|u_j|^2\sin(\omega_j(\tau'-\tau)),
\end{equation}
and the time evolution operator differs only by a time-dependent phase from the one obtained by casting the operators in their natural order
\begin{equation}
\hat{U}(T,0)=e^{i\varphi (T)}\exp\left( -i \int^T_0 d\tau  \hat{H}_{\text{int}} (\tau)  \right),
\end{equation}
where the phase 
\begin{equation}
\varphi (T)=\frac{i}{2}\int^T_0 d\tau_1 \int^T_0 d\tau_2 \Theta(\tau_1-\tau_2)\left[\hat{H}_{\text{int}} (\tau_1),\hat{H}_{\text{int}} (\tau_2) \right]
\end{equation}
describes the ``additional'' motion due to the time ordering. Performing the integral we have
\begin{equation}
\hat{U}(T,0)=e^{i\varphi (T)}\exp\left(\frac{\sigma_z}{2}\sum_j\alpha_j(T)a_j^{\dagger}-\alpha_j^{*}(T)a_j \right),
\end{equation}
where
\begin{equation}
\alpha_j(T):=2u_j^*\frac{1-e^{i\omega_j T}}{ \omega_j}.
\end{equation}
It is not difficult to see that the above formula has the same form as $I_{\pm,j}$. Since $\sigma_z$ in the interaction picture does not contain $e^{\pm i\Omega \tau}$ term and we do not have the resonance effect, we need to consider the composition of all modes. The density matrix of the whole system can be directly obtained as
\begin{equation}
\hat{U}(T,0)[\rho_S \otimes \rho_E] \hat{U}^{\dagger}(T,0).
\end{equation}
We can conclude that the populations of the qubit are unaffected, and the off-diagonal terms are
\begin{equation}
\langle 1|\rho'_S(T)|0\rangle=\sqrt{p(1-p)}\text{Tr}_E\left( \prod_j D_j(\alpha_j(T))\rho_E \right)
\end{equation}
and
\begin{equation}
\langle 0|\rho'_S(T)|1\rangle=\sqrt{p(1-p)}\text{Tr}_E\left( \prod_j D_j(-\alpha_j(T))\rho_E \right),
\end{equation}
where
\begin{equation}
D_j(\alpha_j)=\exp \left( \alpha_j a^{\dagger}_j-\alpha^*_j a_j\right)
\label{displacement}
\end{equation}
is the displacement operator for the $j$-th field mode, while $\text{Tr}_E\left(D_j(\pm \alpha_j)\rho_E \right)$ is the Wigner characteristic function. Since the thermal state $\rho_E$ is Gaussian, we can easily note that it represents a Gaussian function, which immediately leads to the expression
\begin{align}
\text{Tr}_E\left( \prod_j D_j(\pm \alpha_j(T))\rho_E \right)&=\prod_j \exp\left( -\frac{|\alpha_j|^2}{2}\langle {a_j,a^{\dagger}_j}\rangle \right)
\\ \notag
&= \prod_j \exp\left( -\frac{|\alpha_j|^2}{2}(2\bar{n}_j+1) \right).
\end{align}
We can define the above formula as the suppression factor $\chi(\tau)$. On the other hand, the reduced density matrix of the environment is 
\begin{align}
\rho'_E&=(1-p)\prod_j D_j\left(-\frac{\alpha_j}{2}\right)\rho_E D_j^{\dagger}\left(-\frac{\alpha_j}{2}\right)
\\ \notag
&+p\prod_j D_j\left(\frac{\alpha_j}{2}\right)\rho_E D_j^{\dagger}\left(\frac{\alpha_j}{2}\right).
\end{align}
For each $j$, the quantity $\rho_E$ can be described as a distribution in the basis of number state $|n_j\rangle$. Thus $D_j\left(\frac{\alpha_j}{2}\right)\rho_E D_j^{\dagger}\left(\frac{\alpha_j}{2}\right)$ is in fact the same distribution in the basis of the displaced number state $D_j\left(\frac{\alpha_j}{2}\right)|n_j\rangle$ \cite{Oliveira1990}. Using the Baker-Campbell-Haussdorff formula $e^{\hat{B}}\hat{A}e^{-\hat{B}}=\hat{A}+[\hat{B},\hat{A}]+\frac{1}{2}[\hat{B},[\hat{B},\hat{A}]]+\cdots$ and the definition of displacement operator \eqref{displacement}, we have
\begin{align}
D_j^{\dagger}\left(\pm \frac{\alpha_j}{2}\right) a^{\dag}_ja_j D_j\left(\pm\frac{\alpha_j}{2}\right)&=a^{\dag}_ja_j\pm\frac{1}{2}\alpha^*_ja_j \\ \nonumber
&\pm \frac{1}{2}\alpha_ja^{\dagger}_j+\frac{1}{4}|\alpha_j|^2,
\end{align}
thus we can obtain the energy of the environment as
\begin{align}
\text{Tr}\left(\hat{H}_E \rho'_E\right)=\sum_j \left(\bar{n}_j+\frac{|\alpha_j|^2}{4}\right)\omega_j.
\end{align}
and the energy change of the environment as
\begin{equation}
\Delta Q :=\text{Tr}\left[\hat{H}_E(\rho'_E-\rho_E) \right]=\sum_j \frac{|\alpha_j|^2}{4}\omega_j,
\end{equation}
which is always non-negative. On the other hand, since the initial state of the qubit is pure, the difference $\Delta S := S(\rho_S)-S(\rho'_S)$ is always non-positive, thus Landauer's principle is again naturally satisfied. 
 
\begin{figure}
\begin{center}
\includegraphics[width=0.43\textwidth]{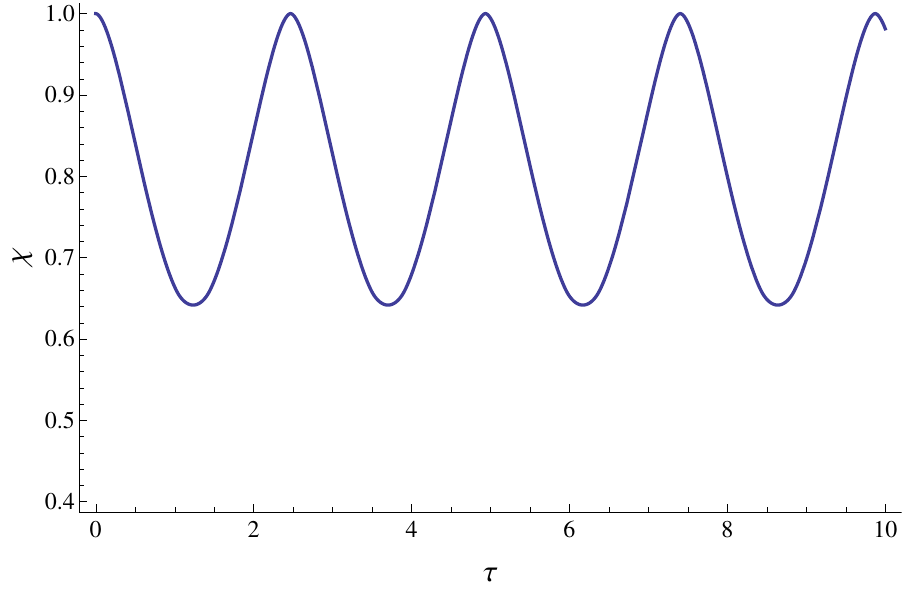}
\caption{The suppression factor $\chi(\tau)$. We choose $p = 0.2$, $L = 1.234$, $x = 0.52345$, $T_E = 1$. }
\label{fig2}
\end{center}
\end{figure}

\begin{figure}
\begin{center}
\includegraphics[width=0.43\textwidth]{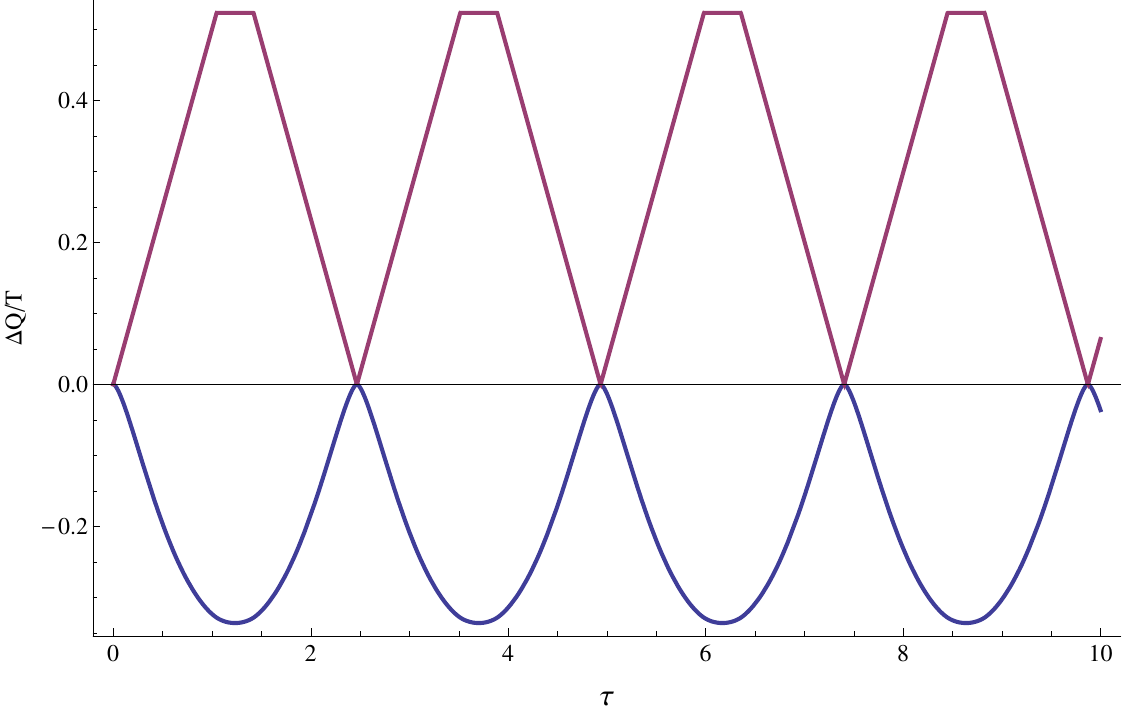}
\caption{From top to bottom the curves correspond to ${\Delta Q}/{T_E}$ and $\Delta S$ respectively. We choose $p = 0.2$, $L = 1.234$, $x = 0.52345$, $T_E = 1$.}
\label{fig3}
\end{center}
\end{figure}

In FIG.\ref{fig2} we present an example of the suppression factor $\chi(\tau)$. Since we are considering discrete models and $|\alpha_j|^2$ has periodicity, we can observe the phenomenon of Poincaré recurrences. Similarly from FIG.\ref{fig3} we can also observe that ${\Delta Q}/{T_E}$ is always non-negative while $\Delta S$ is non-positive, and Poincaré recurrences are also present.

\section{Conclusion}
In the present work we extend the analysis of Landauer's principle in qubit-cavity QFT interaction \cite{Xu:2021buk} to the case that includes decoherence. We analyzed the decoherence of the qubit in the evolutionary process in two cases: $m=\sigma_x$ and $m=\sigma_z$, corresponding to with and without energy dissipation respectively. The initial state of the qubit is choosen to be $|\psi_0\rangle=\sqrt{1-p}|0\rangle+\sqrt{p}|1\rangle$, while the cavity QFT is thermal. The initial density matrix  is different from the mixed density matrix $(1-p)|0\rangle\langle 0|+p|1\rangle\langle 1|$ since it contains coherence (the off-diagonal terms), while the latter may be interpreted as classical probabilities. 

In the $m=\sigma_x$ case (similarly for the $m=\sigma_y$ case), we find the energy change of the cavity QFT and population change of the qubit is identical to the case of $(1-p)|0\rangle\langle 0|+p|1\rangle\langle 1|$. The decoherence process corresponds to a new variation of the off-diagonal terms. The energy change must be accompanied by the decoherence to maintain unitarity. Furthermore, since the initial state of the qubit is pure, the interaction can increase the von Neumann entropy more than in the $(1-p)|0\rangle\langle 0|+p|1\rangle\langle 1|$ case, which has a smaller $\Delta S$. Since Landauer's principle has been proved in the mixed case, it also holds naturally in our case.

In the $m=\sigma_z$ case, we observe that $[\hat{H}_S,\hat{H}_{\text{int}}]= 0$, so the qubit does not undergo energy changes. Since the commutator of $\hat{H}_{\text{int}}$ at different times is just a c-number, the time evolution operator differs only by a time-dependent phase from the one obtained by casting the operators in their natural order. We can directly calculate the evolution of the whole system and perform the partial trace. We verify that the populations of the qubit does remain unchanged, while its off-diagonal terms have a suppression factor. The cavity QFT, on the other hand, becomes the distribution in the basis of the displaced number state. The difference $\Delta S$ is always non-positive, since the initial state is pure, while $\Delta Q$ is always non-negative, thus Landauer's principle is naturally satisfied. The phenomenon of Poincaré recurrences can also be observed.

In \cite{Xu:2021buk} the qubit in accelerated motion (Unruh effect) was also studied. Since the contribution of the acceleration comes from the $|I_{-,j}|^2$ term, this does not affect any of our conclusions. Recently decoherence was proposed as a mean to detect the Unruh effect \cite{Nesterov2020}. We believe that the analysis of qubit-QFT interaction and decoherence can be applied to more contexts, both theoretical and experimental, to deepen our understanding of quantum phenomena. If we use harmonic oscillators instead of qubit to be the system, the interaction can also be solved non-perturbatively in the framework of Gaussian quantum mechanics \cite{Adesso2007,Adesso2014,Xu:2021ihm}. 

We previously mentioned that QFT formulation can describe all fundamental interactions except gravity. However, gravity can also induce decoherence either via classical or quantum gravitational fluctuations, although a clear understanding of gravitational decoherence is still lacking. See, e.g., \cite{Bassi:2017szd,Anastopoulos:2021jdz} for reviews and \cite{Petruzziello2022} for recent development in this area. Future space-based experiments could help to further understand gravitational decoherence and possibly checking the predictions of different quantum theories \cite{2111.05441}.
It is worth remarking that, just as classical thermodynamics is intimately related to classical gravity (e.g., Jacobson derived Einstein field equations assuming the Bekenstein-Hawking entropy and the first law of thermodynamics \cite{J}), quantum information is also expected to play important roles in quantum gravity \cite{2203.07117}. In fact, recently it was discovered that black holes would eventually decohere any quantum superpositions \cite{2205.06279}. Recent studies also included the storage of quantum information at finite temperature in holographic conformal field theories \cite{Banerjee2022}. As Landauer's principle is one of the bridges between thermodynamics and quantum information, understanding it better in more realistic quantum interactions, such as those involving decoherence, is a first tiny step towards the grand goal of quantum gravity. We hope to pursue these gravitational connections in our future research.

Finally, let us mention that in the present work we have only considered the original bound of Landauer's principle: $\Delta Q\geqslant T_E\Delta S$. If we add more conditions to the previous four assumptions mentioned in the Introduction, we may get a tighter bound \cite{Goold2015,Timpanaro2020,Hashimoto:2022lbb,Vu:2022}. The motivation of the these works, including the original paper of Reeb and Wolf \cite{Reeb2013}, are to give some estimates of the evolutionary process of a certain class of systems, i.e. the corresponding bound. These are more general and applicable to a wider range of discussions, allowing us to have a general idea of system evolution without knowing all detailed information about the whole system. In the present work, however, the system is uniquely determined, so that in principle the corresponding physical quantities can be determined directly. We can directly calculate $\Delta Q$ and $\Delta S$ and thus verify Landauer's principle.

\begin{acknowledgments}
Hao Xu thanks Yuan Sun and Fei Meng for useful discussions. He also thanks the Natural Science Foundation of the Jiangsu Higher Education Institutions of China (No.20KJD140001) and the National Natural Science Foundation of China (No.12205250) for funding support. Yen Chin Ong thanks the National Natural Science Foundation of China (No.11922508) for funding support. 
\end{acknowledgments}


\begin{thebibliography}{99}

\bibitem{RevModPhys.76.1267}
M.~ Schlosshauer, ``Decoherence, the Measurement Problem, and Interpretations of Quantum Mechanics'', {\hypersetup{urlcolor=vividviolet}\href{https://journals.aps.org/rmp/abstract/10.1103/RevModPhys.76.1267}{Rev. Mod. Phys. \textbf{76} (2005) 1267}}. 

\bibitem{Zeh1970}
H. D. Zeh, ``On the Interpretation of Measurement in Quantum Theory'', {\hypersetup{urlcolor=vividviolet}\href{https://link.springer.com/article/10.1007/BF00708656}{Found.
Phys. \textbf{1} (1970) 69.}}.

\bibitem{Zurek1981}
W. H. Zurek,  ``Pointer Basis of Quantum Apparatus: Into What Mixture Does the Wave Packet Collapse?'', {\hypersetup{urlcolor=vividviolet}\href{https://journals.aps.org/prd/abstract/10.1103/PhysRevD.24.1516}{Phys. Rev. D \textbf{24} (1981) 1516}}.

\bibitem{Zurek1982}
W. H. Zurek,  ``Environment-Induced Superselection Rules'', {\hypersetup{urlcolor=vividviolet}\href{https://journals.aps.org/prd/abstract/10.1103/PhysRevD.26.1862}{Phys. Rev. D \textbf{26} (1982) 1862}}.

\bibitem{Zurek2003}
W. H. Zurek,  ``Decoherence and the Transition From Quantum to Classical -- REVISITED'', {\hypersetup{urlcolor=vividviolet}\href{https://arxiv.org/abs/quant-ph/0306072}{[arXiv:quant-ph/0306072]}}.

\bibitem{Breuer}
H. Breuer and F. Petruccione,  ``The Theory of Open Quantum Systems'', {Oxford University Press, Oxford,
2002}.

\bibitem{Schlosshauer}
M. Schlosshauer,  ``Decoherence: and the Quantum-to-Classical Transition'', {Springer Science \& Business Media,
2007}.

\bibitem{landauer1961}
R. Landauer, ``Irreversibility and Heat Generation in the Computing Process'', {\hypersetup{urlcolor=vividviolet}\href{https://ieeexplore.ieee.org/document/5392446}{IBM J. Res. Dev. \textbf{5} (1961) 183}}.

\bibitem{landauer1996}
R. Landauer, ``The Physical Nature of Information'', {\hypersetup{urlcolor=vividviolet}\href{https://www.sciencedirect.com/science/article/abs/pii/0375960196004537}{Phys. Lett. A \textbf{217} (1996) 188}}.

\bibitem{Reeb2013}
D. Reeb, M. M. Wolf, ``An Improved Landauer Principle With Finite-Size Corrections'', {\hypersetup{urlcolor=vividviolet}\href{https://iopscience.iop.org/article/10.1088/1367-2630/16/10/103011}{New J. Phys. \textbf{16} (2014) 103011}}.

\bibitem{Olivares2012}
S. Olivares, ``Quantum Optics in the Phase Space'', {\hypersetup{urlcolor=vividviolet}\href{https://link.springer.com/article/10.1140/epjst/e2012-01532-4}{Eur. Phys. J. Special Topics \textbf{203} (2012) 3}}.

\bibitem{1209.4948}
E. Mart\'\i{}n-Mart\'\i{}nez, D. Aasen and A. Kempf, ``Processing Quantum Information with Relativistic Motion of Atoms'', {\hypersetup{urlcolor=vividviolet}\href{https://journals.aps.org/prl/abstract/10.1103/PhysRevLett.110.160501}{Phys. Rev. Lett. \textbf{110} (2013) 160501}}.

\bibitem{Smith2008}
S. T. Smith, R. Onofrio, ``Thermalization in Open Classical Systems With Finite Heat Baths'', {\hypersetup{urlcolor=vividviolet}\href{https://link.springer.com/article/10.1140/epjb/e2008-00070-8}{Eur. Phys. J. B \textbf{61} (2008) 271}}.

\bibitem{Xu:2021buk}
H.~Xu, Y.~C.~Ong and M.~H.~Yung, ``Landauer's Principle in Qubit-Cavity Quantum-Field-Theory Interaction In Vacuum and Thermal States'', {\hypersetup{urlcolor=vividviolet}\href{https://journals.aps.org/pra/abstract/10.1103/PhysRevA.105.012430}{Phys. Rev. A \textbf{105}, no.1 (2022) 012430}}.


\bibitem{Hornberger2009}
K. Hornberger, ``Introduction to Decoherence Theory'', {\hypersetup{urlcolor=vividviolet}\href{https://link.springer.com/book/10.1007/978-3-540-88169-8?noAccess=true}{Lect. Notes Phys. \textbf{768} (2009) 221}}.

\bibitem{Oliveira1990}
F. A. M. de Oliveira, M. S. Kim, P. L. Knight, and V. Buek, ``Properties of Displaced Number States'', {\hypersetup{urlcolor=vividviolet}\href{https://journals.aps.org/pra/abstract/10.1103/PhysRevA.41.2645}{Phys. Rev. A \textbf{41} (1900) 2645 }}.

\bibitem{Nesterov2020}
A. Nesterov, G. Berman, M. Fernández, X. Wang, ``Decoherence as Detector of the Unruh Effect'', {\hypersetup{urlcolor=vividviolet}\href{https://journals.aps.org/prresearch/abstract/10.1103/PhysRevResearch.2.043230}{Phys. Rev. Res. \textbf{2}, no.4 (2020) 043230}}.

\bibitem{Adesso2007}
G. Adesso and F. Illuminati, ``Entanglement in Continuous-Variable Systems: Recent Advances and Current Perspectives'', {\hypersetup{urlcolor=vividviolet}\href{https://iopscience.iop.org/article/10.1088/1751-8113/40/28/S01}{J. Phys. A: Math. Theor. \textbf{40} (2007) 7821}}.

\bibitem{Adesso2014}
G. Adesso, S. Ragy, A. R. Lee, ``Continuous Variable Quantum Information: Gaussian States and Beyond'', {\hypersetup{urlcolor=vividviolet}\href{https://www.worldscientific.com/doi/abs/10.1142/S1230161214400010}{Open Syst. Inf. Dyn. \textbf{21} (2014) 1440001}}.

\bibitem{Xu:2021ihm}
H.~Xu and S.~Y.~Chen,
``Entropy Production and Correlation Spreading in the Interaction Between Particle Detector and Thermal Baths'',
{\hypersetup{urlcolor=vividviolet}\href{https://arxiv.org/abs/2111.04050}{[arXiv:2111.04050 [quant-ph]]}}.


\bibitem{Bassi:2017szd}
A.~Bassi, A.~Gro\ss{}ardt and H.~Ulbricht,
``Gravitational Decoherence'',
{\hypersetup{urlcolor=vividviolet}\href{https://iopscience.iop.org/article/10.1088/1361-6382/aa864f}{Class. Quant. Grav. \textbf{34}, no.19 (2017) 193002}}.

\bibitem{Anastopoulos:2021jdz}
C.~Anastopoulos and B.~L.~Hu,
``Gravitational Decoherence: A Thematic Overview'', {\hypersetup{urlcolor=vividviolet}\href{https://avs.scitation.org/doi/abs/10.1116/5.0077536}{AVS Quantum Sci. \textbf{4}, no.1 (2022) 015602}}.

\bibitem{Petruzziello2022}
L. Petruzziello and F. Illuminati,
``Quantum Gravitational Decoherence From Fluctuating Minimal Length and Deformation Parameter At the Planck Scale'', {\hypersetup{urlcolor=vividviolet}\href{https://www.nature.com/articles/s41467-021-24711-7}{Nature Communications volume \textbf{12}, Article number: 4449 (2021)}}.

\bibitem{2111.05441}
C.~Anastopoulos, M.~Blencowe and B.~L.~Hu, ``Gravitational Decoherence in Deep Space Experiments'', {\hypersetup{urlcolor=vividviolet}\href{https://arxiv.org/abs/2111.05441}{[arXiv:2111.05441 [gr-qc]]}}.


\bibitem{J}
T. Jacobson, ``Thermodynamics of Spacetime: The Einstein Equation of State'', {\hypersetup{urlcolor=vividviolet}\href{https://journals.aps.org/prl/abstract/10.1103/PhysRevLett.75.1260}{Phys. Rev. Lett. \textbf{75} (1995) 1260}}.

\bibitem{2203.07117}
T. Faulkner, T. Hartman, M. Headrick, M. Rangamani, B. Swingle, ``Snowmass White Paper: Quantum Information in Quantum Field Theory and Quantum Gravity'', {\hypersetup{urlcolor=vividviolet}\href{https://arxiv.org/abs/2203.07117}{[arXiv:2203.07117 [hep-th]]}}.


\bibitem{2205.06279}
D. L. Danielson, G. Satishchandran, R. M. Wald, ``Black Holes Decohere Quantum Superpositions'', {\hypersetup{urlcolor=vividviolet}\href{https://arxiv.org/abs/2205.06279}{[arXiv:2205.06279 [hep-th]]}.}

\bibitem{Banerjee2022}
A. Banerjee, T. Kibe, N. Mittal, A. Mukhopadhyay, and P. Roy,
 ``Erasure Tolerant Quantum Memory and the Quantum Null Energy Condition in Holographic Systems'', {\hypersetup{urlcolor=vividviolet}\href{https://journals.aps.org/prl/abstract/10.1103/PhysRevLett.129.191601}{Phys. Rev. Lett. \textbf{129} (2022) 191601}}.

\bibitem{Goold2015}
J. Goold, M. Paternostro, K. Modi, ``Nonequilibrium Quantum Landauer Principle'', {\hypersetup{urlcolor=vividviolet}\href{https://journals.aps.org/prl/abstract/10.1103/PhysRevLett.114.060602}{Phys. Rev. Lett. \textbf{114} (2015) 060602}}.

\bibitem{Timpanaro2020}
A. M. Timpanaro, J. P. Santos, G. T. Landi, ``Landauer's Principle at Zero Temperature'', {\hypersetup{urlcolor=vividviolet}\href{https://journals.aps.org/prl/abstract/10.1103/PhysRevLett.124.240601}{Phys. Rev. Lett. \textbf{124} (2020) 240601}}.

\bibitem{Hashimoto:2022lbb}
K.~Hashimoto and C.~Uchiyama,
``Effect of Quantum Coherence on Landauer\textquoteright{}s Principle,''
{\hypersetup{urlcolor=vividviolet}\href{https://www.mdpi.com/1099-4300/24/4/548}{Entropy \textbf{24}, no.4 (2022) 548}}.

\bibitem{Vu:2022}
T. V. Vu and K. Saito
``Finite-Time Quantum Landauer Principle and Quantum Coherence,''
{\hypersetup{urlcolor=vividviolet}\href{https://journals.aps.org/prl/abstract/10.1103/PhysRevLett.128.010602}{Phys. Rev. Lett. \textbf{128},(2022) 010602}}.


\end{thebibliography}
\end{document}